\pdfoutput=1
\documentclass[journal=acscii,manuscript=article]{achemso}
\setkeys{acs}{usetitle=true}
\usepackage{setspace}
\usepackage[utf8]{inputenc}
\usepackage{amsmath}
\usepackage{graphicx}
\usepackage{subcaption}
\usepackage{float}
\usepackage[font=small]{caption}
\usepackage[font=small]{subcaption}
\usepackage[usenames, dvipsnames]{color}

\newcommand{\edit}[1] { \textcolor{black}{#1}}

\title{Rapid Prediction of Electron-Ionization Mass Spectrometry using Neural Networks}

\author{Jennifer N. Wei}
\email{weijennifer@google.com}
\affiliation{Google Brain, Cambridge MA 02142, USA}
\alsoaffiliation{Department of Chemistry and Chemical Biology, Harvard University, Cambridge MA 02138, USA}

\author{David Belanger}
\email{dbelanger@google.com}
\affiliation{Google Brain, Cambridge MA 02142, USA}

\author{Ryan P. Adams}
\affiliation{Departmment of Computer Science, Princeton University, Princeton NJ 08540, USA}

\author{D. Sculley}
\affiliation{Google Brain, Cambridge MA 02142, USA}

\begin{document}

\maketitle

\section*{Abstract}
When confronted with a substance of unknown identity, researchers often perform mass spectrometry on the sample and compare the observed spectrum to a library of previously-collected spectra to identify the molecule. While popular, this approach will fail to identify molecules that are not in the existing library. 
In response, we propose to improve the library’s coverage by augmenting it with synthetic spectra that are predicted \edit{from candidate molecules} using machine learning. 
We contribute a lightweight neural network model that quickly predicts mass spectra for small molecules, averaging 5 ms per molecule with a recall-at-10 accuracy of 91.8\%. Achieving high-accuracy predictions requires a novel neural network architecture that is designed to capture typical fragmentation patterns from electron ionization. We analyze the effects of our modeling innovations on library matching performance and compare our models to prior machine learning-based work on spectrum prediction.

\section{Introduction}

Mass spectrometry (MS) is an important tool used to identify unknown molecular samples in a variety of applications, from characterization of organic synthesis products, to pharmacokinetic studies \cite{massspec_pharmakinetics}, to forensic studies \cite{Zhou2017LatentFingerprints}, to analyzing gaseous samples on remote satellites \cite{Petrie_ions_in_space}.
In electron-ionization mass spectrometry (EI-MS), molecular samples are ionized by an electron beam and broken into fragments. The resultant ions are separated by an electric field until they reach a detector. The mass spectrum is a distribution of the frequency or intensity of each type of ion, ordered by mass-to-charge (\textit{m/z}) ratio.

A popular method for identifying a sample from its mass spectrum is to look up the sample's spectrum in a \textit{reference library}~\cite{2017nist,mclafferty2016wiley}. Here, a similarity function is used to measure the similarity between the query spectrum from the sample and each spectrum in the library. If the measurement noise when obtaining the query spectrum is reasonable, then the library spectrum with the highest similarity will correspond to the correct identification of the sample~\cite{stein1995ChemicalSubstructureIdentification,stein1994optimization}. A schematic of this process is shown in Figure~\ref{fig:library_matching}a.

This library matching approach is very popular, but it suffers from a \textit{coverage problem}: if the sample consists of a molecule that is not in the library, then correct identification is impossible. This is an issue in practice, since existing mass spectral reference libraries, such as the NIST/NIH/EPA MS database~\cite{2017nist}, Wiley Registry of Mass Spectral Data~\cite{mclafferty2016wiley}, and MassBank~\cite{horai2010massbank} only contain hundreds of thousands of reference spectra. The coverage problem could be reduced by recording spectra for additional molecules, but this is time consuming and expensive. For example, NIST releases updates to its library every 3 years, containing roughly 20,000 new spectra. Additionally, mass spectra of new molecules are only added to the library if the molecule is of common interest; molecules for newly synthesized compounds are typically not incorporated~ \cite{2017nist,stein2012MassLibReview}.

An alternative solution is to use \textit{de novo} methods that input a spectrum and directly generate a molecule, without using a fixed list of molecules; we discuss some of these methods in the Background section. 
These approaches currently have low-accuracy and are difficult for practitioners to incorporate into their existing work-flows.

Another method for alleviating the coverage problem is to augment existing libraries with synthetic spectra that are generated by a model. Thus far, this approach has not been practical, as existing spectrum prediction methods are very computationally expensive. These prediction models use quantum mechanics calculations ~\cite{bauer2016compute,grimme2013towards,Guerra_BEB_model} or machine learning~\cite{allen2016computational} to estimate the probability of each bond breaking under ionization, and thus the frequency of each ion fragment. Since these methods must either compute molecular orbital energies with high accuracy using expensive calculations, or else stochastically simulate the fragmentation of the molecule, the time needed for each model to make a prediction scales with the size of the molecule, taking up to 10 min for large molecules \cite{bauer2016compute,allen2016computational}. For applications of identifying metabolites from a metabolomic spectra, much faster predictions of individual molecular spectra are required~\cite{VIANT201764}.

In response, we present Neural Electron Ionization Mass Spectrometry (NEIMS), a neural network that predicts the electron-ionization mass spectrum for a given small molecule. Since our model directly predicts spectra, instead of bond breaking probabilities, it is dramatically faster than previously reported methods, making it possible to generate predictions for thousands of possible candidates in seconds. Furthermore, the approach does not rely on specific details of EI, and thus our model could be easily retrained to predict mass spectra for other ionization methods. \edit{We envision that this tool can expand coverage in areas of molecular space of interest to researchers that are likely candidates for the identity of the molecule.}

We test the performance of our model by predicting mass spectra for small molecules from the NIST 2017 Mass Spectral Library. We find that the predictive capability of our model is similar to previously reported machine learning models, but requires much less time to make predictions. Additionally, we report the similarity of the spectra predicted by NEIMS. The code repository for NEIMS is publicly available at github.com/brain-research/deep-molecular-massspec.

\section{Background}
\label{sec:related-work}

\subsection{De novo prediction of molecules from mass spectra}
Several algorithms have been developed previously for either predicting spectra or for predicting the molecule's identity given the spectrum. One of the earliest efforts in artificial intelligence was a model used to identify molecules from their mass spectrum. Heuristic DENDRAL (Dentritic Algorithm) was a collaboration between chemists and computer scientists at Stanford in the 1960s~\cite{buchanan1981dendral}. This algorithm used expert rules from chemistry to help identify patterns in the spectra and suggest possible identities for the molecule. A few years later, Meta DENDRAL was introduced to learn the expert rules that originally been given to Heuristic DENDRAL~\cite{lindsay1993dendral}.
    
Since then, several models have been reported to predict identities of samples directly from the spectrum.
Many have been developed for tandem mass spectrometry, where the task is to predict the original peptide sequences from digested fragments given the mass spectrum~\cite{Eng1994sequest}. Some of these methods use machine learning to achieve this task~\cite{Tran8247deepnovo, Schoenholtz2018supervision}.
Several previously published models use neural network models to analyze mass spectra to predict molecule directly. One approach predicts the fingerprint from the spectrum and looks up the fingerprint in a library of fingerprints~\cite{duhrkop2015searching}. Another approach predicts the molecule directly either as a SMILES representation~\cite{curry1990msnet,spec2smiles} or from a ranked list of possible structural conformers~\cite{Lim2018ChemicalStructure}. 

\subsection{Prediction of mass spectra from molecules}
In this work, we focus on the prediction of spectra from molecules. The advantage of this approach over de novo approaches is that new libraries of synthetic spectra can be easily incorporated into the existing mass spectrometry software to improve the coverage of existing libraries.
  
The first prediction methods for EI-MS spectrum used quantum mechanical simulation techniques to predict fragmentation events. There are three methods of predicting the mass spectrum using first principles~\cite{bauer2016compute}.
The first is to use quasi-equilibrium theory, also known as Rice-Ramsberger-Kassel-Marcus theory, to estimate the rate constants for ionization reaction~\cite{lorquet1994whither, lorquet2000landmarks, rosenstock1952absolute}.
The second is to estimate the bond order energies within a molecule, and estimate where a molecule may fragment. A related method to this second method is to calculate the cross-section of molecular orbitals upon electron impact to predict the molecule's ionization behavior~\cite{irikura2017ab, Guerra_BEB_model}.
The third method uses Born-Oppenheimer Molecular Dynamics. Quantum Chemistry Electron-Ionization Mass Spectrometry (QCEIMS) is a particularly recent example of the ab initio molecular dynamics method~\cite{grimme2013towards,Asgeirsson_QCEIMS,bauer2016compute}.
The trajectories resulting from this simulation are then analyzed for the presence of ionic fragments. The distribution of the ion fragments aggregated from all the simulations is then renormalized to generate a calculated EI-MS spectrum.
Each of these methods requires at least 1000 seconds per molecule \cite{allen2016computational}, and may even take days or weeks for molecules of 50 atoms. While these methods may be fast for methods involving density functional theory, they do not have the speed needed to rapidly generate a collection of spectra thousands of molecules. Furthermore, some of the basis sets used for the density functional theory might not support the presence of inorganic atoms.
    
Allen et al.~\cite{allen2016computational} introduced a machine learning model, Competitive Fragmentation Modeling-Electron Ioinization (CFM-EI), to predict EI-MS spectra. This probabilistic model predicts the probability of breaking molecular bonds under electron ionization, and also predicts the charged fragment that is likely to form. In order to generate the spectra, it is necessary to run a stochastic simulation to determine the frequency of each molecular fragment.
In results section, 
we directly compare this method with our proposed model.

\section{Methods}

\begin{figure}
    \centering
    \begin{subfigure}{0.9\textwidth}
        \includegraphics[width=\textwidth]{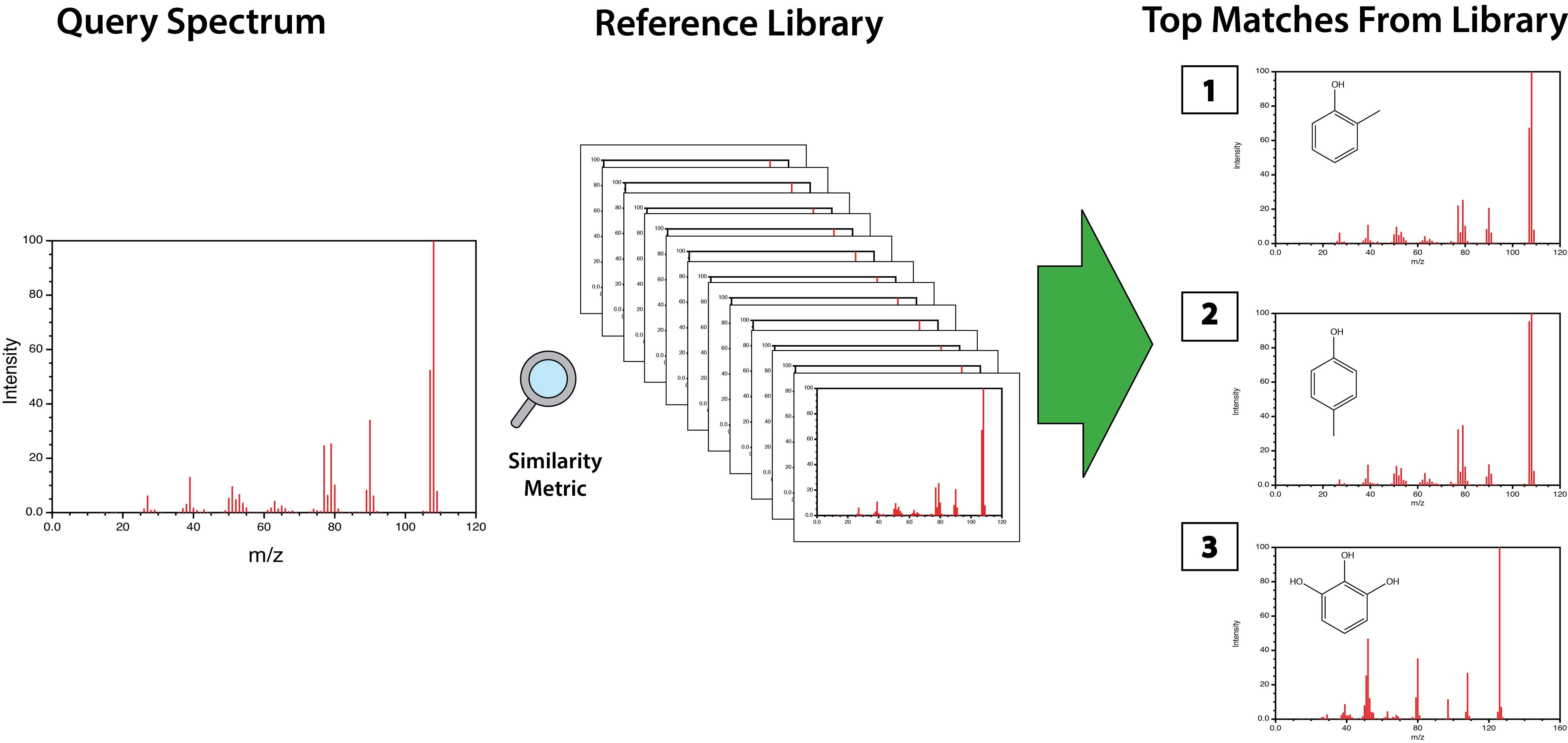}
        \caption{}
    \end{subfigure}
    \begin{subfigure}{0.9\textwidth}
        \includegraphics[width=\textwidth]{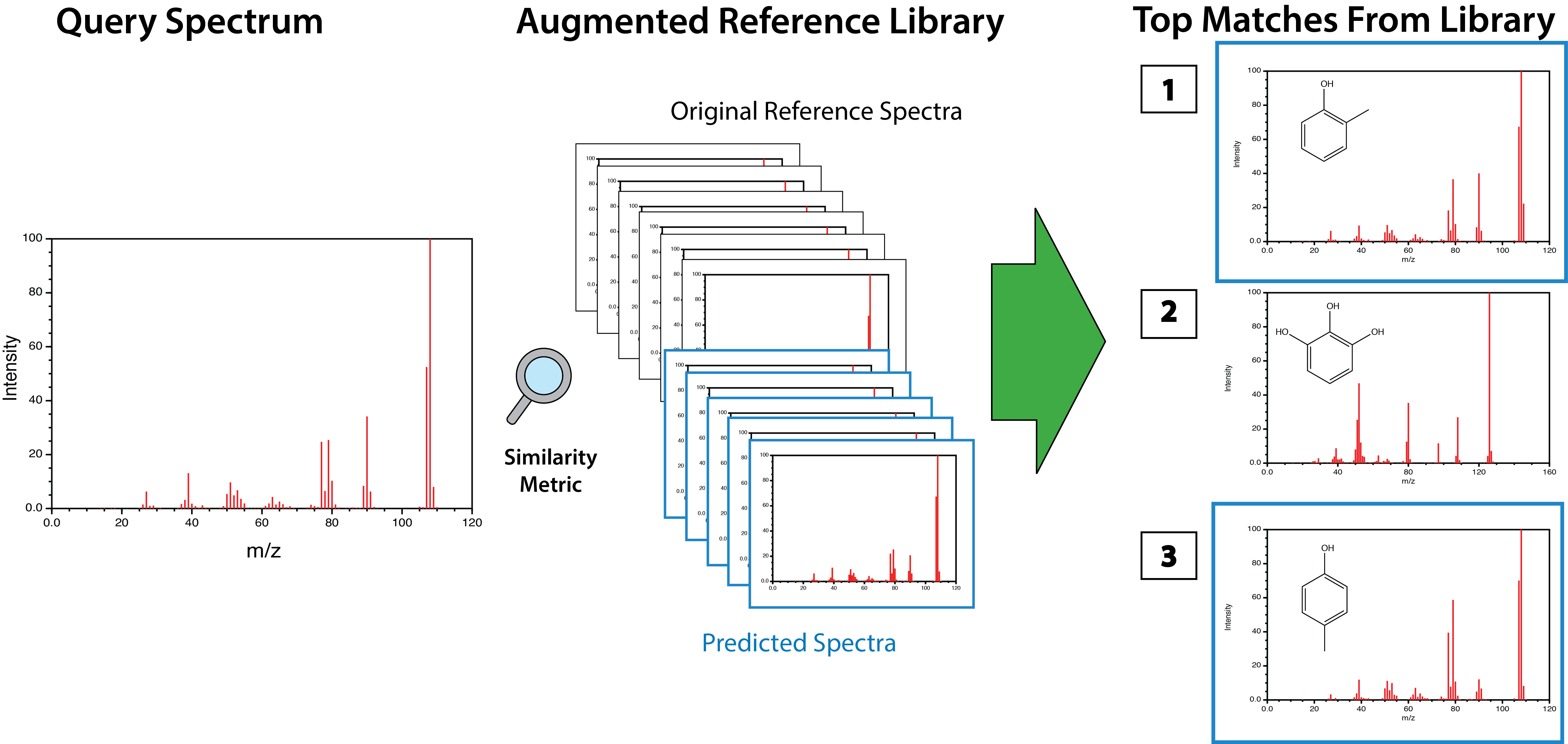}
        \caption{}
    \end{subfigure}
    \caption[Library Matching Task]{Library matching task.~(a) Depiction of how query spectra are matched to a collection of reference spectra as performed by mass spectrometry software. (b) Query spectra are compared against a library comprising of  spectra from the NIST 2017 main library and spectra predicted by our model (outlined in blue).  Spectral images adapted from NIST Webbook~\cite{NIST_WebBook}.}
    \label{fig:library_matching}
\end{figure}

Our goal is to design a model that will accurately predict the EI-MS spectrum for any molecule. This will be used to produce an augmented reference library containing both predicted spectra and experimentally-measured spectra.  This task is outlined in Figure \ref{fig:library_matching}b. 

We first discuss our choice of similarity metrics for mass spectra. Next, we describe our method for spectra prediction, describing our new architecture which is required for high accuracy of mass spectra. We then explain how we evaluate our model's impact on the library matching task.

\subsection{Similarity Metrics for Mass Spectra}\label{sec:similarity_discussion}

The ability to a match a query spectrum from a sample to the correct spectrum in the library depends on the choice of similarity metric between spectra~\cite{mclafferty1974probability,stein1994optimization}.
A weighted cosine similarity is commonly used by mass spectrometry software. The exact form of the cosine similarity is given below~\cite{stein1994optimization}:

\begin{equation}\label{eq:stein-similarity}
\text{Similarity}(\boldsymbol{I}_q, \boldsymbol{I}_l) = \frac{\sum_{k=1}^{M_{max}} m_{k} I_{qk}^{0.5} \cdot m_{k} I_{lk}^{0.5}}{\left\lVert\sum_{k=1}^{M_q} (m_{k} I_{qk}^{0.5})^2\right\rVert \left\lVert\sum_{k=1}^{M_l} (m_{k} I_{lk}^{0.5})^2\right\rVert}.
\end{equation}
Here, {$\boldsymbol{I}_q$} and {$\boldsymbol{I}_l$} are vectors of \textit{m/z} intensities representing the query spectrum and the library spectrum respectively, $m_k$ and $I_k$ are the mass-to-charge ratio and intensity found at $m/z = k$, $M_l$ and $M_q$ are the largest indices of $\boldsymbol{I}_q$ and $\boldsymbol{I}_l$ with nonzero values,and $M_{max}$ is the larger of $M_l$ and $M_q$. The motivation for the weighting by $m/z$ is because the peaks in mass spectra corresponding to larger fragments are more characteristic and useful in practice for identifying the true molecule.

Other similarity metrics besides cosine distance similarities are also employed. For example, one other similarity method involves estimating the relative importance of one peak given the other peaks~\cite{mclafferty1974probability}. Other methods uses a Euclidian difference between peaks, or use a variation of the Hamming distance~\cite{stein1994optimization,hertz1971identification}. Another similarity metric accounts for neutral losses, or the intensity peaks corresponding to the loss of small, neutral fragments from the original molecular ion~\cite{moorthy2017combining}. It is also possible to use the same form of the similarity function as in \eqref{eq:stein-similarity}, but with different weighting given to the intensity or the masses~\cite{stein1994optimization}. In principle, machine learning could be also used to learn a parametrized similarity metric that yields improved library matching performance. However, this custom metric would be difficult to deploy, since it would require changing the software used by practitioners. 

We develop our model with the assumption that Eq.~\eqref{eq:stein-similarity} will be used for the similarity metric in downstream library matching software that consumes an augmented library.

\subsection{Spectral Prediction}\label{sec:ms_spectral_prediction}

We treat the prediction of mass spectrometry spectra as a multidimensional regression task. The output of our model is a vector that represents the intensity at every integral \textit{m/z} bin. We use this discretization granularity for \textit{m/z} because it is what is provided in the NIST data sets we use for training our model.

In the NEIMS model (Figure~\ref{fig:model_prediction}), we first map molecules to additive Extended Circular Fingerprints (ECFPs)~\cite{rdkit}.  
These fingerprints are similar to their binary counterparts~\cite{Rogers_2010_ECFP} in that they record molecular subgraphs made up from local neighborhoods around each atom node in the molecule, but differ in that they count the occurrences for each subgroup. This information is then hashed into a vector representation. The difference is that additive fingerprints record the frequency that each bit is set, rather than just the presence. The RDKit Cheminformatics package~\cite{rdkit} was used to generate the fingerprints. We use a fingerprint length of 4096 with a radius of 2.
These features are then passed into a multilayer perceptron neural network (MLP). 
To account for some of the physical phenomena of ionization, we make some application-specific adjustments to the prediction from the MLP, described in the 'Adjustments for Physical Phenomena' section. 

In the Library Matching Results section, 
we compare the performance of NEIMS to that of a simple linear regression (LR) model. Here, we apply a linear transformation to the ECFP features.

To train the model, we use a modified mean-squared-error loss function. This loss function, shown below, follows the same weighting pattern as in Eq.~\ref{eq:stein-similarity}:

\begin{equation}\label{eq:training_loss}
    L(\boldsymbol{I}, \hat{\boldsymbol{I}}) = \sum_{k=1}^{M(x)} \left(\frac{m_k I_k^{0.5}}{\left\lVert\sum_{k=1}^{M} (m_{k} I_{k}^{0.5})^2\right\rVert} - \frac{m_k \hat{I_k}^{0.5}}{\left\lVert\sum_{k=1}^{M} (m_{k} \hat{I}_{k}^{0.5})^2\right\rVert}\right)^2
\end{equation}

where $I$ is the ground truth spectrum, $\hat{I}$ is the predicted spectrum, and $M(x)$ is the mass of the input molecule. 
We used stochastic gradient descent to optimize the parameters of the MLP with the Adam optimizer~\cite{Kingma_adam_optimizer}. We use Tensorflow~\cite{Tensorflow-2016} to construct and train the model.

\subsection{Adjustments for Physical Phenomena}\label{sec:ms_physical_phenomena}

\begin{figure}
    \centering
    \includegraphics[width=0.9\linewidth]{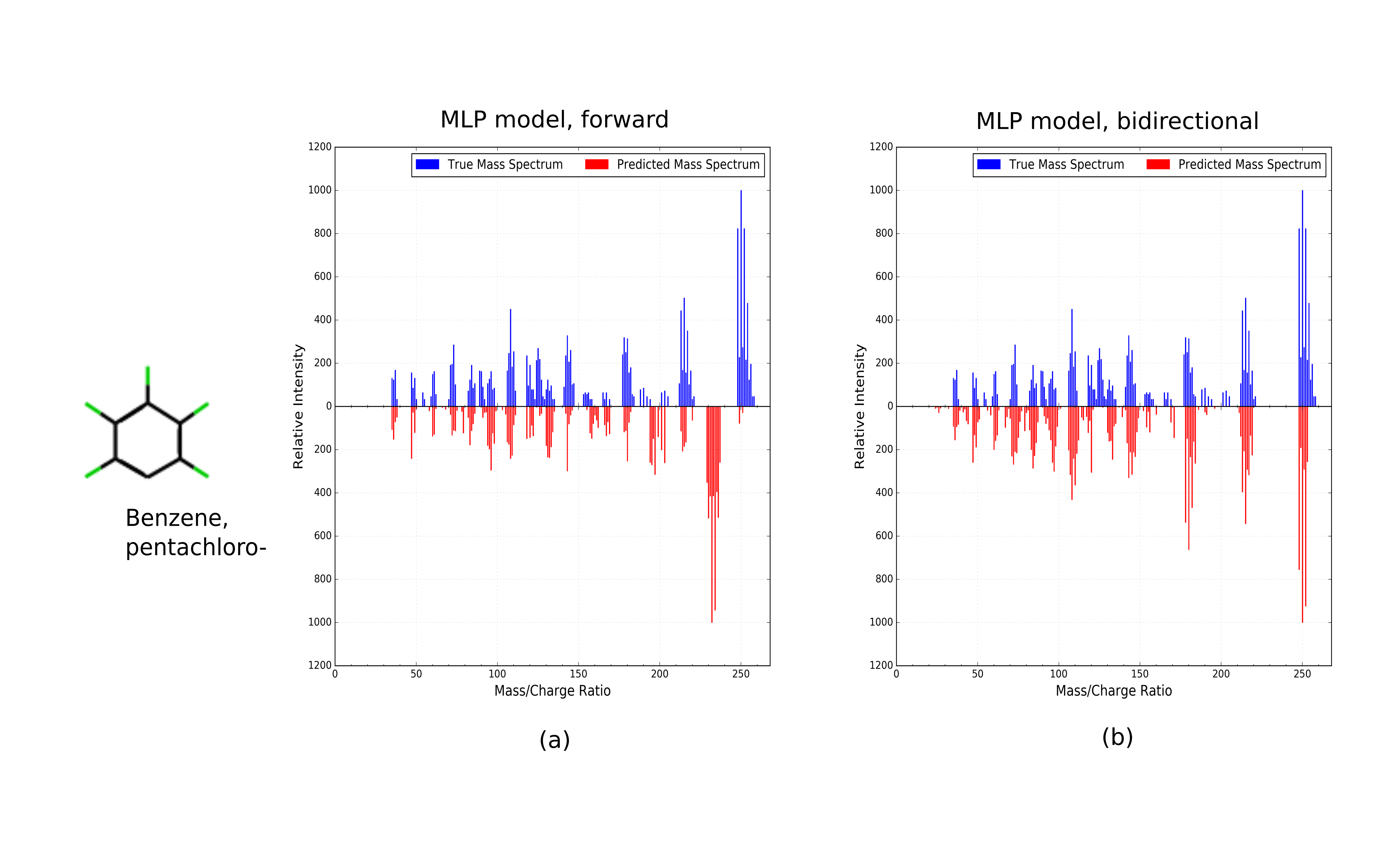}
        \caption[Sample Spectra Prediction]{Spectral prediction with MLP forward model (a) and MLP bidirectional model (b). For both spectra plots, the true spectrum is shown in blue on top, while the predicted spectrum is shown inverted in red. ~Note that the spectrum predicted by the bidirectional model shows fewer stray peaks than the forward model, particularly for larger \textit{m/z} values. These peaks are much easier to predict with the reverse prediction mode.}
    \label{fig:MLP_improvement_spectra}
\end{figure}

In practice, we have found that the conventional MLP described in the previous section struggles to accurately predict the right-hand side of spectra (Figure~\ref{fig:MLP_improvement_spectra}a). Errors in this region, which correspond to large $m/z$, are particularly damaging for library matching with the weighting in~\eqref{eq:stein-similarity}.

\begin{figure}
    \centering
    \includegraphics[width=0.9\linewidth]{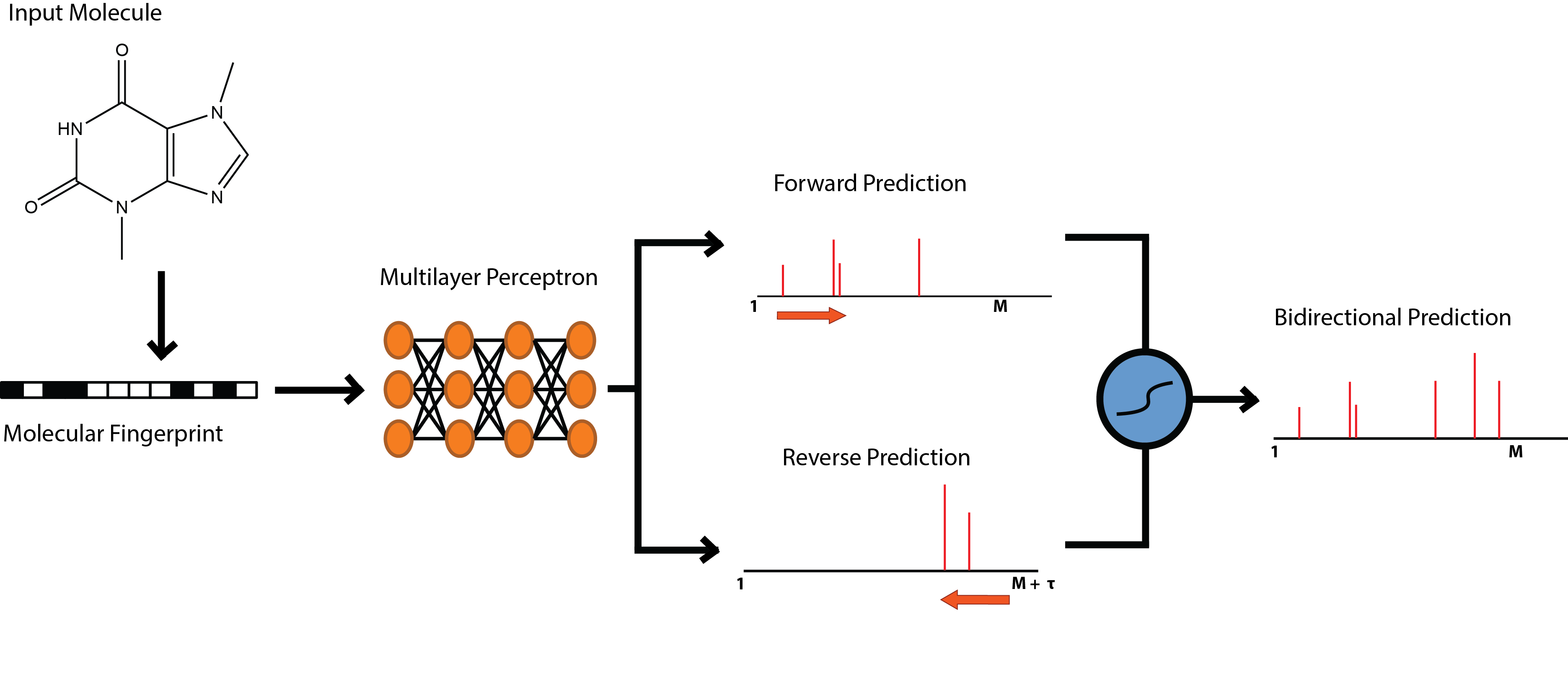}
    \caption[Neural Electron Ionization MS Prediction Model]{Molecular representations are passed into a multilayer perceptron to generate an initial output. This output is used to make a forward prediction starting at $\textit{m/z}=0$ and $\textit{m/z}=M$ and in reverse starting from $\textit{m/z}=M$ and ending at $\textit{m/z}=0$. A sigmoid gating is applied to the inputs as shown in Eq. \ref{eq:sigmoid_gate}}
    \label{fig:model_prediction}
\end{figure}

This section introduces a revised neural network architecture (Figure~\ref{fig:model_prediction}) designed to better model the underlying fragmentation process that occurs in mass spectrometry. We have found that it improves prediction in the high mass region of the spectrum (Figure~\ref{fig:MLP_improvement_spectra}b), which yields improvements in library matching as discussed in the Results section. 

As is standard for MLPs used for regression, the predictions of the above MLP model on an input molecule $x$ are an affine transformation of a set of features $f(x)$, which are computed by all but the final layer of the network. For reasons that will become apparent, we refer to the above MLP as performing \textit{forward} prediction. At bin $m/z = i$, we have the following predicted intensity:
\begin{equation}
p^f_i(x) = w^{f\top}_i f(x) + b^f_i, \label{eq:pred-i}
\end{equation}
where $w^f_i$ and $b^f_i$ are the model's weights and biases for forward prediction at bin $i$. 

The input ECFP features, from which $f(x)$ is computed, capture local structures in the molecule, so generally $f(x)$ will be more accurate in capturing the presence of small substructures of molecule $x$. Often, there is a direct correspondence between the presence of such substructures and spectral peaks with small $m/z$. For example, in Figure \ref{fig:MLP_improvement_spectra}a a peak occurs at $m/z = 35$, due to the presence of chlorine. Therefore an accurate forward prediction model will have a learned weight $w_{35}$ that will output a high intensity at $i = 35$ if there is evidence in $f(x)$ for the presence of chlorine. 

On the other hand, forward prediction often struggles to accurately predict intensities for large fragments that are the result of neutral losses~\cite{stein1995ChemicalSubstructureIdentification}. One reason for this is that the composition of large fragments is not captured well by the ECFP representation. Another reason is that information learned about the cleavage of a small group does not transfer well across molecules of different masses. For pentachlorobenzene, which has a molecular mass of 250 Da, the fragment that results from the loss of a neutral chlorine atom results in a peak at 215 Da. Meanwhile, for chlorobenzene, which has a mass of 112 Da, the fragment resulting from a loss of a chlorine atom would have a peak at 77 Da. Despite the clear relationship between these intensity peaks, the forward model is not parametrized to capture this pattern.

In response, following the physical phenomenon that created the fragments, we define larger ion peaks as a function of the residual groups that were broken off from the original molecule. 
Referring to our previous example of pentachlorobenzene ($M(x) = 250$), we can parametrize the $m/z$ ratio of the fragment which lost a chlorine group as $m/z = 250 - 35 = 215$. The corresponding fragment in chlorobenzene would have a mass of $m/z = 215 - 35 =  77$. By defining the peaks in this way, it is possible for these predictions of spectral intensities to be linked by the prediction at index 35.
This leads to the indexing scheme of our \textit{reverse prediction} model:
\begin{equation}
p^r_{M(x) + \tau - i}(x) = w^{r\top}_i f(x) + b^r_i, \label{eq:pred-i-reverse}
\end{equation}
Here, $\tau > 0$ is a small shift that allows for peaks to occur at intensities greater than $M(x)$, due to isotopes. In practice, reverse prediction is implemented using a copy of the forward model, with separate sets of parameters for the final affine layer, but shared parameters for $f(x)$. The outputs of this model are postprocessed on a per-molecule basis to obey the indexing in~\eqref{eq:pred-i-reverse}, which depends on each molecule's mass. 

Both the forward and reverse predictions are combined to form a \textit{bidirectional} prediction. That is, the final prediction at index $i$ is a combination of both $p^f_i$ and $p^r_i$. In the case of pentachlorobenzene, the prediction of spectral intensity at $m/z = 215$ is a function of $p^f_{215}$ from the forward mode and $p^r_{35 + \tau}$ from the reverse mode. Instead of simply averaging the two prediction modes, we have found that small additional performance improvements can be obtained using a coordinate-wise \textit{gate}. Here, the output $p_i(x)$ at position $i$ is given by:
\begin{equation}\label{eq:sigmoid_gate}
p_i(x) = \sigma(\text{gate}_i) ~ p^f_i(x) + (1 - \sigma(\text{gate}_i)) ~ p^r_{i}(x),
\end{equation}
where $\text{gate}_i$ is an affine transformation of $f(x)$ and $\sigma(\cdot)$ is a sigmoid function. This approach echoes the formulation of the Hybrid Similarity Search designed by Moorthy et al.~\cite{moorthy2017combining}, which accounts for peaks that are created by small fragment ions and those which are created by large fragments which have lost smaller groups.

Finally, for all models, we zero out predicted intensities at $m/z$ that are greater than $M(x) + \tau$.

By adding these features, we incorporate some of the physical phenomena that occur in mass spectrometry into our model while maintaining the overall simplicity of the MLP. In this way, we are able to predict the spectrum directly without resorting to sampling bond-breaking events within the molecule, which requires subsequent stochastic sampling to obtain a spectrum.

\subsection{Library Matching Evaluation}\label{sec:library_matching_description}

We evaluate NEIMS using an \textit{augmented reference library} consisting of a combination of observed spectra and model-predicted spectra, with library matching performance computed with respect to a \textit{query set} of spectra. These are from the NIST 2017 replicates library, which is a collection of noisier spectra for molecules that are contained in the NIST main library. The inconsistencies in these spectra reflect experimental variation, and make an informative data set to test our model's performance.

To construct the augmented reference library, we edit the NIST main library, removing spectra corresponding to the query set molecules and replacing them with the predictions from NEIMS. 
We then perform library matching and calculate the similarity between each query spectrum and every spectrum from the augmented library. We record the rank of the correct spectrum, i.e. the rank of the predicted spectrum corresponding to the molecule which made the query spectrum. The similarity metric is Eq.~\eqref{eq:stein-similarity}.

For the purposes of tuning model hyperparameters, we chose to optimize recall@10, i.e. the percentage of our query set for which the correct spectra had a matching rank of less than or equal to 10 in the library matching task. Half of the replicates library was used for tuning hyperparameters, and the remaining half was used to evaluate test performance. All models were trained on the spectra prediction task for 100,000 training steps with a batch size of 100.

During the library match search, we have a \textit{mass filtering} option. This feature reduces the library size so it only includes spectra from molecular candidates that have a molecular mass that differ by a few Daltons from the mass of the query molecule.
If the EI-MS analysis is combined with mass spectrometry techniques using soft ionization methods, it is possible to determine the mass of the molecule being analyzed. In the CFM-EI model, the molecular formula is used to filter the search library
~\cite{allen2016computational}. Using the molecular mass to filter the library allows more possible candidate spectra to be considered in the search than using a molecular formula filter.

\section{Results and Discussion}

To analyze the performance of the models, we trained with 240,942 spectra from the NIST 2017 Mass Spectral Main Library. These spectra were selected so that no molecules in the replicates library have spectra in the training set.

After hyperparameter tuning using Vizier~\cite{Google_Vizier}, we found that the optimal MLP architecture has seven layers of 2000 nodes, with residual network connections between the layers~\cite{he_resnet}, using ReLU activation and a dropout rate of 0.25. 

\subsection{Library Matching Results}
\label{sec:library-matching-results}
\begin{figure}
    \centering
    \begin{subfigure}[b]{0.52\linewidth}
        \includegraphics[width=\linewidth]{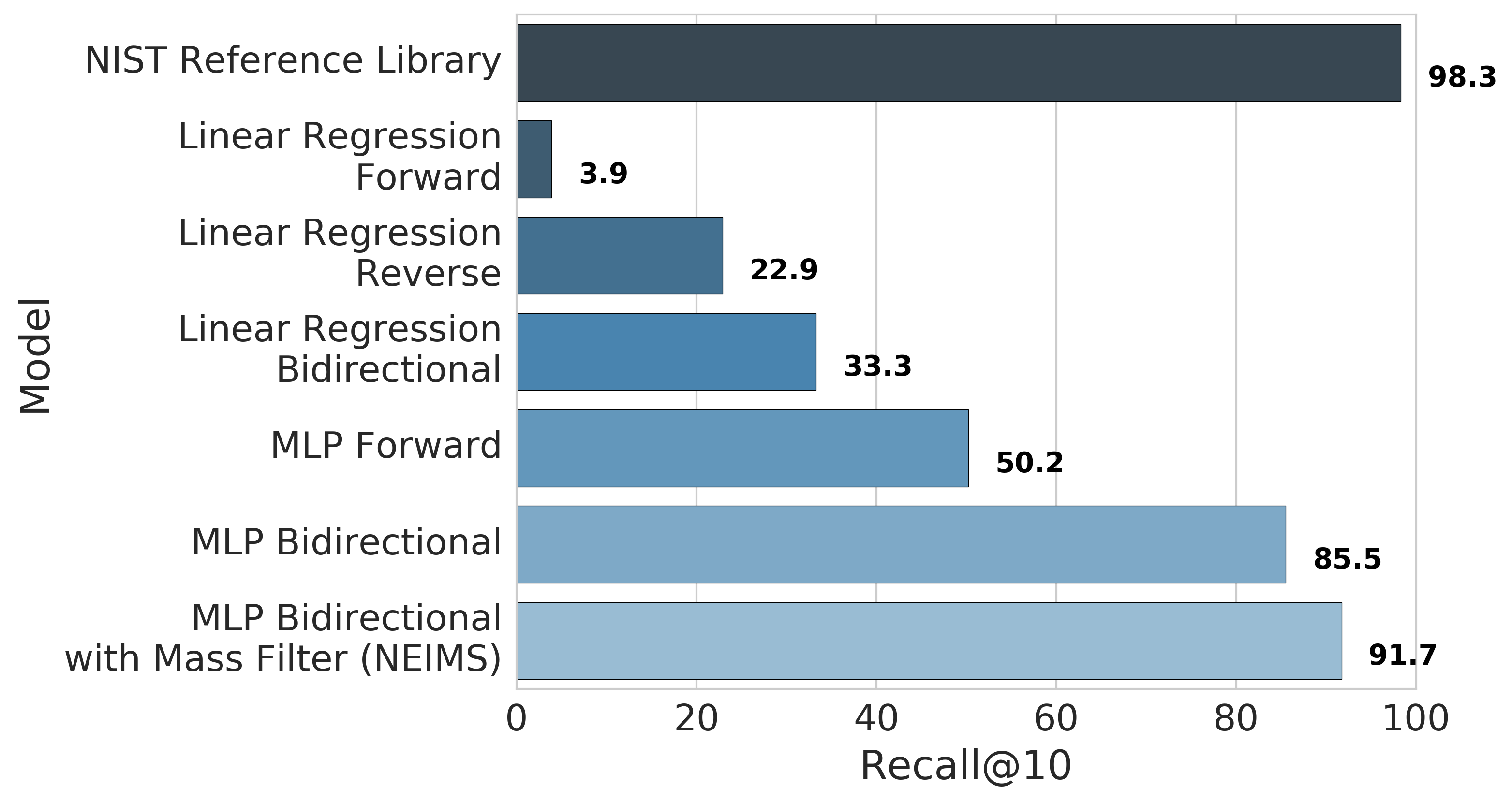}
        \caption[Library Matching Results]{}
    \end{subfigure}
    \begin{subfigure}[b]{0.42\linewidth}
        \includegraphics[width=\linewidth]{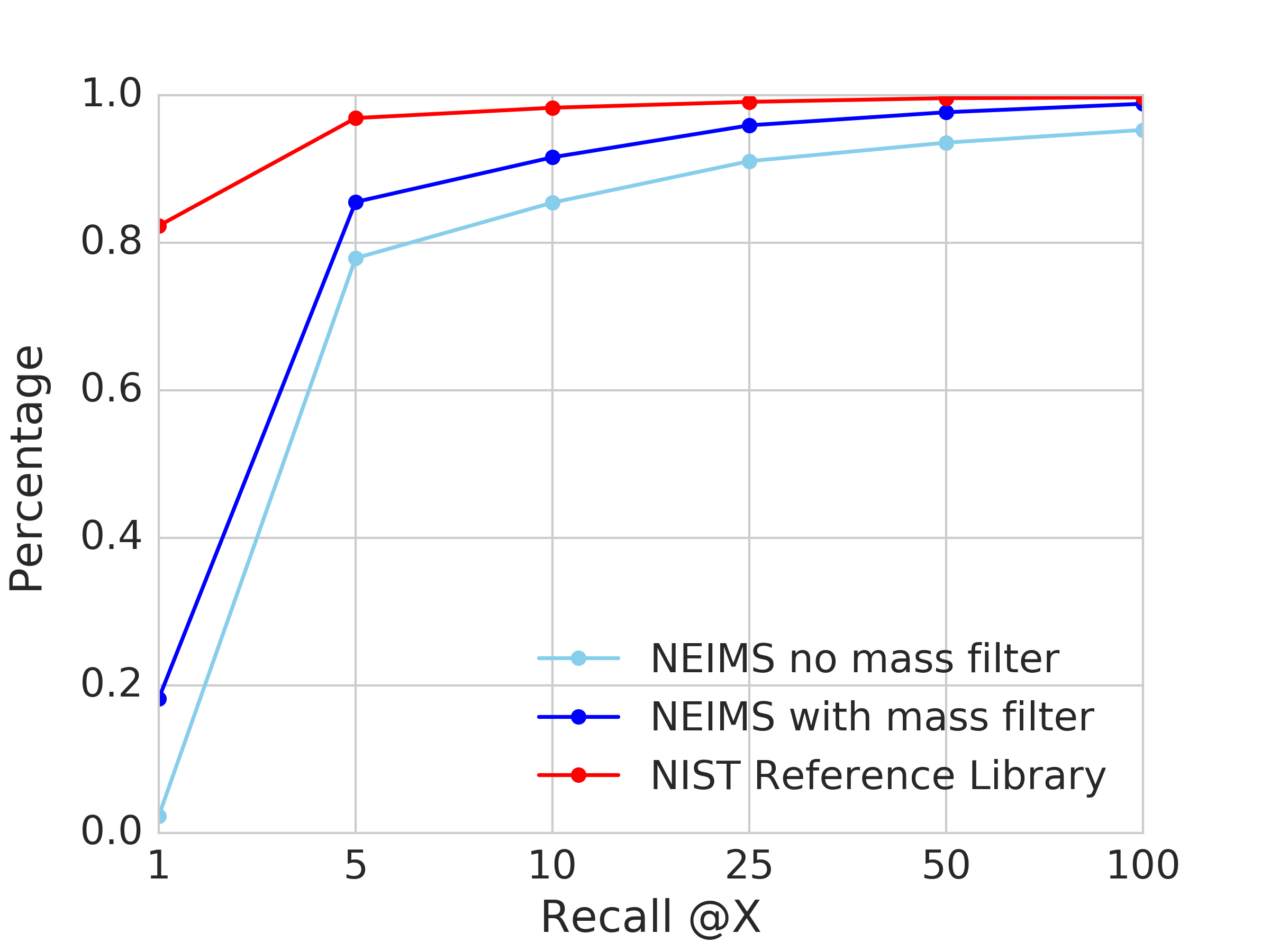}
        \caption[Recall@X Results for Different models]{}
    \end{subfigure}
    \caption{Performance of different model architectures. \edit{Figure a) compares the recall@10 accuracy of the linear regression model and the multi-layer perceptron model using the forward, reverse, and bidirectional architecture. Figure b) shows the performance of NEIMS at different recall levels, and compares it against the performance of using the NIST main library itself.}}
    \label{fig:main_results_and_recall_results}
\end{figure}

We first examine the effects of our various modeling decisions on performance. Figure \ref{fig:main_results_and_recall_results}a compares the performance of forward, reverse, and bidirectional versions of the linear regression and MLP models on the library matching task. For bidirectional prediction in the linear regression model, the forward and reverse predictions are simply averaged together, rather than applying the gate described in \eqref{eq:sigmoid_gate}.

The top row of Figure \ref{fig:main_results_and_recall_results}a shows that it is not possible to achieve perfect recall accuracy on the library matching task even when using the full NIST main library as the reference library, without any model-predicted spectra. Observing Figure \ref{fig:main_results_and_recall_results}b we see that using the NIST main library as the reference library, we have 86\% recall@1 accuracy, and 98.3\% recall@10 accuracy. This serves as a practical upper bound on achievable library matching accuracy and reflects the experimental inconsistencies between between the main library spectra and replicates spectra~\cite{stein2012MassLibReview}.

The forward prediction mode for both the linear regression model and the multilayer perceptron (MLP) has poor performance. The linear regression model is improved by ~20\% when switching to using  reverse mode prediction. Using  bidirectional prediction mode improves  recall@10 accuracy by 30\% for both the linear regression and the multilayer perceptron model. This finding suggests that the bidirectional prediction mode is more effective at capturing the fragmentation events than the forward-only model.

Figure \ref{fig:MLP_improvement_spectra} shows the improvement in spectral prediction for pentachlorobenzene using the bidirectional MLP model. Note that the bidirectional model on the right more accurately models intensities at larger \textit{m/z}. The intensity peaks for larger \textit{m/z} are critical for determining the identity of a molecule, and are more heavily weighted in Eq.~\eqref{eq:stein-similarity}.

NEIMS achieves 91.7\% recall@10 after applying a mass filter. The mass filter was set to a tolerance of 5 Da of the query molecule's mass; this reduces the size of the library to a median of 6,696 spectra for each query molecule. In practice, this tolerance window could be set to a larger window, depending on the uncertainty of the information about the molecular mass of the ion.
For the rest of this report, we will refer to the bidirectional multilayer perceptron model with mass filtering of 5 Daltons as the default settings for NEIMS.

From Figure \ref{fig:main_results_and_recall_results}b we see that while NEIMS has decent performance for recall levels of 10 and above compared to the NIST spectral library, it has considerably worse performance for recall values of 1 and 5. This result is unsurprising given that the hyperparameters of the model were trained to maximize performance on recall@10. \edit{As many experimentalists will examine the top matches to find the library spectrum that best matches the sample's spectrum, we believe that tuning for recall@10 is sufficient for an initial approach. }

\subsection{Comparison to previously reported models}
\label{sec:NIST14-results}

We next compared our model's performance directly to the performance of the CFM-EI model~\cite{allen2016computational}. The setup of Allen et al. differs from our current setup in a few ways. First, they evaluate their model on the NIST '14 spectral library. Second, for the library matching task, their augmented reference library contains only spectra predicted by their model, and none from the original NIST collection. Third, the cosine similarity metric Eq. \eqref{eq:stein-similarity} used for evaluation in library matching in CFM-EI uses a different weighting scheme. In their analysis, the cosine similarity is weighted by $m_k^{0.5}$ instead of $m_k$ in order to de-emphasize the larger peaks in the mass spectrum, as they ran their experiments on other data sets with a higher proportion of larger molecules~\cite{allen2016computational}.

To compare the performance of NEIMS to that of CFM-EI, we match their setup identically. We retrain our NEIMS model on the NIST 14 data set, and evaluate the performance using the NIST 14 replicates as the query set. For library matching, we incorporate only predicted spectra into our augmented library, and using the same modified similarity metric.

\begin{table}
    \centering
    \begin{tabular}{cccc}
        Model & Recall@1 (\%) & Recall@10 (\%) &  Average run time (ms) \\
        \hline
        NIST '14 Reference Library & 77 & 99* & -- \\
        CFM-EI & 42.6 & 89* &  300,000 \\
        NEIMS & 54.3 & 92.7 &  0.47
    \end{tabular}
    \caption{Performance on Library matching task for NIST 17. * indicates that values were estimated from  Figure 4 of Allen et al.~\cite{allen2016computational}}
    \label{tab:NIST14_results}
\end{table}

The library matching performance for CFM-EI and NEIMS are compared against the NIST14 library for library matching performance are reported in Table \ref{tab:NIST14_results}. NEIMS performs slightly better than CFM-EI on the library matching task. More importantly, NEIMS is able to make spectral predictions orders of magnitude faster than CFM-EI.  With NEIMS, it would be possible to generate spectra for 1 million molecules in 90 min on a CPU, with potential for considerable speedup with using GPU.

\subsection{Distances between predicted and ground truth spectra}

\begin{figure}
    \centering
    \includegraphics[scale=0.4]{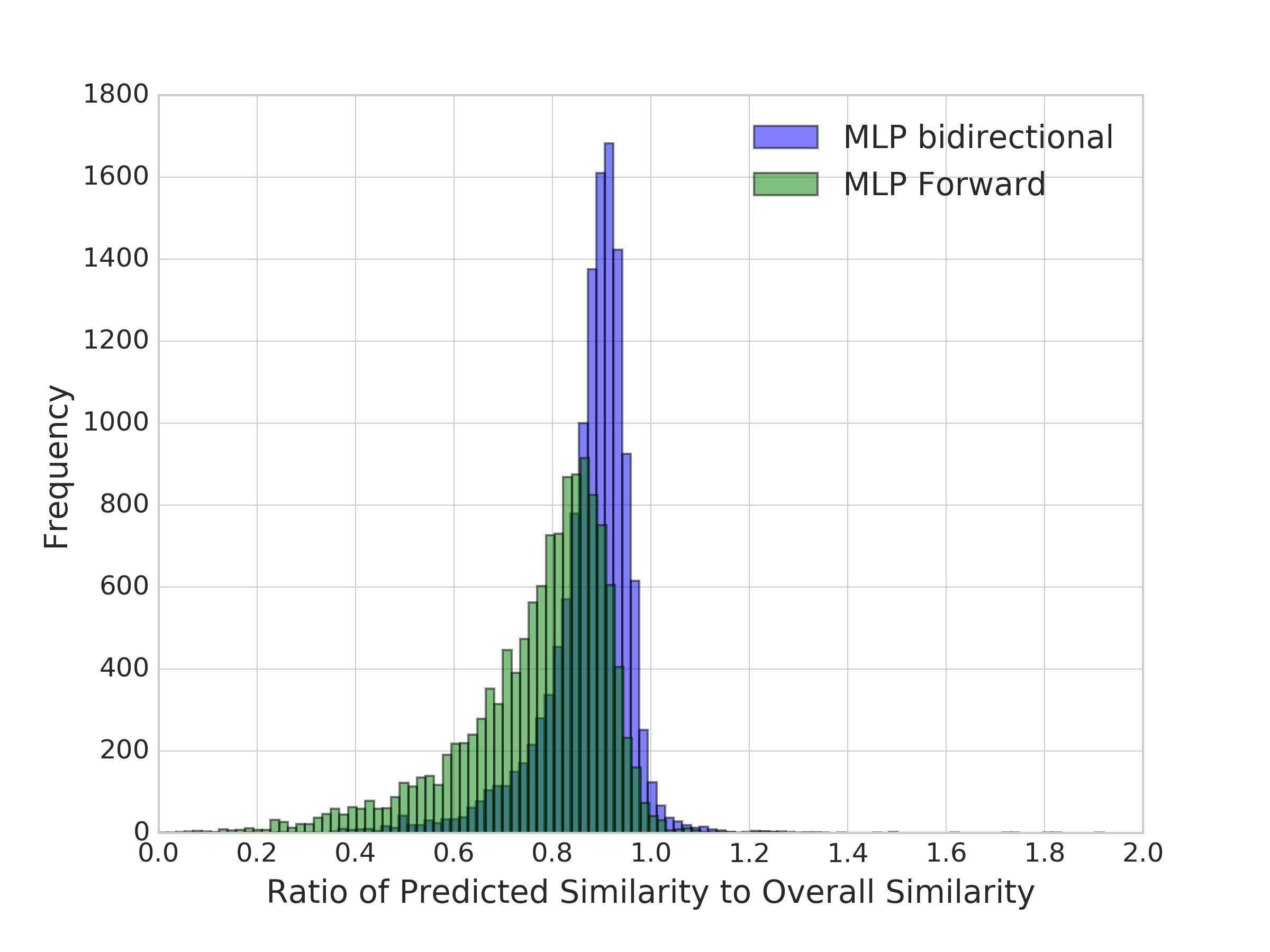}
        \caption[Similarity Analysis of NEIMS predicted spectra to spectra self-similarity]{Comparing the similarity between the predicted spectrum and the ground truth spectrum to the overall similarity between spectra for the same molecule. }
    \label{fig:similarity_analysis}
\end{figure}

So far, we have evaluated the quality of the NEIMS predictions indirectly, by way of how they affect library matching with an augmented library. Next, we assess the prediction accuracy directly, by measuring the similarity (Eq.~\ref{eq:stein-similarity}) between spectra in the NIST main library and the model's predictions. We refer to this similarity as the \textit{predicted similarity}.

There is inherent noise in mass spectra due to stochasticity of the underlying physical process and also to experimental inconsistencies~\cite{stein2012MassLibReview}. The NIST replicates library provides multiple spectra for each molecule, and we can use these sets of spectra to characterize the scale of this noise for each molecule. Specifically, we define the inherent noise for a given molecule as the average pairwise similarity between all corresponding spectra, in both the NIST main library and the NIST replicates library, and refer to this as the \textit{overall similarity}. 

For each molecule, we compute the ratio of the predicted similarity to overall similarity as a normalized metric for the quality of our predictions. A ratio of 1.0 would suggest that there are is limited available headroom for improvements using machine learning, since the model's errors are comparable to the variability in the data.

Figure \ref{fig:similarity_analysis} shows the improvement in this ratio for the MLP bidirectional model over the MLP forward model, confirming that the bidirectional model has better spectral prediction performance. For the MLP bidirectional model, roughly half of the molecules have a predicted similarity to overall similarity ratio that is greater than 0.9, indicating that there is potential for further improvement to the model. Some of these molecules have ratios that are greater than 1, which is possible if there is more variation between the spectra (i.e. a lower overall similarity) than between the predicted spectrum and the main library spectrum (i.e. predicted similarity).

\edit{To analyze the ability of our model to extrapolate we analyzed the relationship between predicted similarity and similarity of the query molecule to the training set molecules. The molecules in the test set have limited structural similarity to the molecules in the training set. We observe that 22.7\% of molecules in the test set have a Tanimoto similarity of greater than 0.8 with at least one molecule in the training set. A plot of this relationship between predicted spectral similarity and closely related molecules in terms of Tanimoto similarity can be found in  Supplementary Figure 3. Based on these results, we believe that our model is able to extrapolate to some areas of molecular space that were not fully covered by the training set. Future work will examine the limitations of the model's ability to extrapolate.}

\section{Conclusion}

We demonstrate that NEIMS achieves high library matching performance on an augmented spectral library containing predictions for molecules in the query set. 
The performance of NEIMS is also slightly better than existing machine learning models for predicting EI-MS spectra, with significant boost in speed of prediction.

The high performance in library matching is attributable to the bidirectional prediction mode. The reverse mode in particular allows the model to more accurately predict intensities for larger fragments which result from the loss of small neutral subgroups. We observe that the improvement in the library matching task also corresponds with improvement in the similarity of the predicted spectra to the ground truth spectra.

Several adjustments could be made to further improve the predictive accuracy of NEIMS. For example, NEIMS currently does not have a method to model intensity peaks corresponding to isotopes in ion fragments. If we were to train on spectral data with greater precision in the peaks locations, we should be able to learn the exact identities of atoms based on the decimal values of the \textit{m/z} peak locations.

Mass filtering improved the performance of NEIMS by 6\%. This suggests that for experimental setups where it is possible to know the molecular mass of the sample with some accuracy, it is possible to improve the accuracy of matching on the augmented spectral library. It would also be interesting to explore other settings for mass filtering, such as filtering out spectra which have a molecular mass that is much smaller than the position of the largest \textit{m/z} peak.

\edit{Using graph-convolutional molecular representations~\cite{duvenaud_convolutional_2015, gilmer_2017_mpnn}, especially bond-centered representations~\cite{kearnes2016molecular} ,  would likely improve predictive accuracy at a slightly higher computational cost.}
The predictions made from ECFP are limited by the descriptiveness of the fingerprint~\cite{rdkit_blogpost_collide_bits}. In particular, the overlap in representation for different molecular features represents a huge limitation to the representation of the molecule. 

Combining NEIMS with transfer learning methods could allow for spectral prediction specific to individual spectrometry machines. A library of such machine-specific spectra would improve matching~\cite{stein2012MassLibReview}.

The lightweight framework of NEIMS makes it possible to rapidly generate spectral predictions for large numbers of molecular candidates. This collection of predicted spectra can then be used directly in mass spectrometry software to expand the coverage of molecules which can be identified by mass spectrometry. Because the requirements of NEIMS has limited dependence to EI mass spectrometry, it likely that some of the principles used here could be extended to other types of mass spectrometry.

\section{Acknowledgments}
We thank Stephen Stein and Laura Castellanos for fruitful discussions about mass spectrometry and for their helpful feedback on the manuscript. We thank Steven Kearnes for his suggestions on the manuscript, and Lucy Colwell and Michael Brenner for their helpful conversations.

\section{Supplementary information}
The code for this work can be found at github.com/brain-research/deep-molecular-massspec. 

The supplementary information section contains peripheral findings on the relationship between Tanimoto similarity and predicted spectra similarity, as well as the training and test loss curves.

\bibliographystyle{unsrt}
\bibliography{references}

\providecommand{\latin}[1]{#1}
\makeatletter
\providecommand{\doi}
  {\begingroup\let\do\@makeother\dospecials
  \catcode`\{=1 \catcode`\}=2 \doi@aux}
\providecommand{\doi@aux}[1]{\endgroup\texttt{#1}}
\makeatother
\providecommand*\mcitethebibliography{\thebibliography}
\csname @ifundefined\endcsname{endmcitethebibliography}
  {\let\endmcitethebibliography\endthebibliography}{}
\begin{mcitethebibliography}{43}
\providecommand*\natexlab[1]{#1}
\providecommand*\mciteSetBstSublistMode[1]{}
\providecommand*\mciteSetBstMaxWidthForm[2]{}
\providecommand*\mciteBstWouldAddEndPuncttrue
  {\def\EndOfBibitem{\unskip.}}
\providecommand*\mciteBstWouldAddEndPunctfalse
  {\let\EndOfBibitem\relax}
\providecommand*\mciteSetBstMidEndSepPunct[3]{}
\providecommand*\mciteSetBstSublistLabelBeginEnd[3]{}
\providecommand*\EndOfBibitem{}
\mciteSetBstSublistMode{f}
\mciteSetBstMaxWidthForm{subitem}{(\alph{mcitesubitemcount})}
\mciteSetBstSublistLabelBeginEnd
  {\mcitemaxwidthsubitemform\space}
  {\relax}
  {\relax}

\bibitem[Hsieh and Korfmacher(2006)Hsieh, and
  Korfmacher]{massspec_pharmakinetics}
Hsieh,~Y.; Korfmacher,~W.~A. Increasing speed and throughput when using
  HPLC-MS/MS systems for drug metabolism and pharmacokinetic screening.
  \emph{Current Drug Metabolism} \textbf{2006}, \emph{7}, 479--489\relax
\mciteBstWouldAddEndPuncttrue
\mciteSetBstMidEndSepPunct{\mcitedefaultmidpunct}
{\mcitedefaultendpunct}{\mcitedefaultseppunct}\relax
\EndOfBibitem
\bibitem[Zhou and Zare(2017)Zhou, and Zare]{Zhou2017LatentFingerprints}
Zhou,~Z.; Zare,~R.~N. Personal information from latent fingerprints using
  desorption electrospray ionization mass spectrometry and machine learning.
  \emph{Analytical Chemistry} \textbf{2017}, \emph{89}, 1369--1372, PMID:
  28194988\relax
\mciteBstWouldAddEndPuncttrue
\mciteSetBstMidEndSepPunct{\mcitedefaultmidpunct}
{\mcitedefaultendpunct}{\mcitedefaultseppunct}\relax
\EndOfBibitem
\bibitem[Petrie and Bohme(2006)Petrie, and Bohme]{Petrie_ions_in_space}
Petrie,~S.; Bohme,~D.~K. Ions in space. \emph{Mass Spectrometry Reviews}
  \textbf{2006}, \emph{26}, 258--280\relax
\mciteBstWouldAddEndPuncttrue
\mciteSetBstMidEndSepPunct{\mcitedefaultmidpunct}
{\mcitedefaultendpunct}{\mcitedefaultseppunct}\relax
\EndOfBibitem
\bibitem[Stein(2017)]{2017nist}
Stein,~S.~E. National {I}nstitute of {S}tandards and {T}echnology ({NIST})
  {M}ass {S}pectral {D}atabase. 2017\relax
\mciteBstWouldAddEndPuncttrue
\mciteSetBstMidEndSepPunct{\mcitedefaultmidpunct}
{\mcitedefaultendpunct}{\mcitedefaultseppunct}\relax
\EndOfBibitem
\bibitem[Mclafferty(2016)]{mclafferty2016wiley}
Mclafferty,~F.~W. \emph{Wiley Registry of Mass Spectral Data}, 11th ed.; John
  Wiley and Sons, 2016\relax
\mciteBstWouldAddEndPuncttrue
\mciteSetBstMidEndSepPunct{\mcitedefaultmidpunct}
{\mcitedefaultendpunct}{\mcitedefaultseppunct}\relax
\EndOfBibitem
\bibitem[Stein(1995)]{stein1995ChemicalSubstructureIdentification}
Stein,~S.~E. Chemical substructure identification by mass spectral library
  searching. \emph{Journal of the American Society for Mass Spectrometry}
  \textbf{1995}, \emph{6}, 644--655\relax
\mciteBstWouldAddEndPuncttrue
\mciteSetBstMidEndSepPunct{\mcitedefaultmidpunct}
{\mcitedefaultendpunct}{\mcitedefaultseppunct}\relax
\EndOfBibitem
\bibitem[Stein and Scott(1994)Stein, and Scott]{stein1994optimization}
Stein,~S.~E.; Scott,~D.~R. Optimization and testing of mass spectral library
  search algorithms for compound identification. \emph{Journal of the American
  Society for Mass Spectrometry} \textbf{1994}, \emph{5}, 859--866\relax
\mciteBstWouldAddEndPuncttrue
\mciteSetBstMidEndSepPunct{\mcitedefaultmidpunct}
{\mcitedefaultendpunct}{\mcitedefaultseppunct}\relax
\EndOfBibitem
\bibitem[Horai \latin{et~al.}(2010)Horai, Arita, Kanaya, Nihei, Ikeda, Suwa,
  Ojima, Tanaka, Tanaka, Aoshima, \latin{et~al.} others]{horai2010massbank}
others,, \latin{et~al.}  MassBank: A public repository for sharing mass
  spectral data for life sciences. \emph{Journal of mass spectrometry}
  \textbf{2010}, \emph{45}, {703--714}\relax
\mciteBstWouldAddEndPuncttrue
\mciteSetBstMidEndSepPunct{\mcitedefaultmidpunct}
{\mcitedefaultendpunct}{\mcitedefaultseppunct}\relax
\EndOfBibitem
\bibitem[Stein(2012)]{stein2012MassLibReview}
Stein,~S. Mass spectral reference libraries: an ever-expanding resource for
  chemical identification. \emph{Analytical Chemistry} \textbf{2012},
  \emph{84}, 7274 -- 7282\relax
\mciteBstWouldAddEndPuncttrue
\mciteSetBstMidEndSepPunct{\mcitedefaultmidpunct}
{\mcitedefaultendpunct}{\mcitedefaultseppunct}\relax
\EndOfBibitem
\bibitem[Bauer and Grimme(2016)Bauer, and Grimme]{bauer2016compute}
Bauer,~C.~A.; Grimme,~S. How to compute electron ionization mass spectra from
  first principles. \emph{The Journal of Physical Chemistry A} \textbf{2016},
  \emph{120}, 3755--3766\relax
\mciteBstWouldAddEndPuncttrue
\mciteSetBstMidEndSepPunct{\mcitedefaultmidpunct}
{\mcitedefaultendpunct}{\mcitedefaultseppunct}\relax
\EndOfBibitem
\bibitem[Grimme(2013)]{grimme2013towards}
Grimme,~S. Towards first principles calculation of electron impact mass spectra
  of molecules. \emph{Angewandte Chemie International Edition} \textbf{2013},
  \emph{52}, 6306--6312\relax
\mciteBstWouldAddEndPuncttrue
\mciteSetBstMidEndSepPunct{\mcitedefaultmidpunct}
{\mcitedefaultendpunct}{\mcitedefaultseppunct}\relax
\EndOfBibitem
\bibitem[{Guerra} \latin{et~al.}(2012){Guerra}, {Parente}, {Indelicato}, and
  {Santos}]{Guerra_BEB_model}
{Guerra},~M.; {Parente},~F.; {Indelicato},~P.; {Santos},~J.~P. Modified binary
  encounter Bethe model for electron-impact ionization. \emph{International
  Journal of Mass Spectrometry} \textbf{2012}, \emph{313}, 1--7\relax
\mciteBstWouldAddEndPuncttrue
\mciteSetBstMidEndSepPunct{\mcitedefaultmidpunct}
{\mcitedefaultendpunct}{\mcitedefaultseppunct}\relax
\EndOfBibitem
\bibitem[Allen \latin{et~al.}(2016)Allen, Pon, Greiner, and
  Wishart]{allen2016computational}
Allen,~F.; Pon,~A.; Greiner,~R.; Wishart,~D. Computational prediction of
  electron ionization mass spectra to assist in GC/MS compound identification.
  \emph{Analytical chemistry} \textbf{2016}, \emph{88}, 7689--7697\relax
\mciteBstWouldAddEndPuncttrue
\mciteSetBstMidEndSepPunct{\mcitedefaultmidpunct}
{\mcitedefaultendpunct}{\mcitedefaultseppunct}\relax
\EndOfBibitem
\bibitem[Viant \latin{et~al.}(2017)Viant, Kurland, Jones, and
  Dunn]{VIANT201764}
Viant,~M.~R.; Kurland,~I.~J.; Jones,~M.~R.; Dunn,~W.~B. How close are we to
  complete annotation of metabolomes? \emph{Current Opinion in Chemical
  Biology} \textbf{2017}, \emph{36}, 64 -- 69, Omics\relax
\mciteBstWouldAddEndPuncttrue
\mciteSetBstMidEndSepPunct{\mcitedefaultmidpunct}
{\mcitedefaultendpunct}{\mcitedefaultseppunct}\relax
\EndOfBibitem
\bibitem[Buchanan and Feigenbaum(1981)Buchanan, and
  Feigenbaum]{buchanan1981dendral}
Buchanan,~B.~G.; Feigenbaum,~E.~A. \emph{Readings in artificial intelligence};
  Elsevier, 1981; pp 313--322\relax
\mciteBstWouldAddEndPuncttrue
\mciteSetBstMidEndSepPunct{\mcitedefaultmidpunct}
{\mcitedefaultendpunct}{\mcitedefaultseppunct}\relax
\EndOfBibitem
\bibitem[Lindsay \latin{et~al.}(1993)Lindsay, Buchanan, Feigenbaum, and
  Lederberg]{lindsay1993dendral}
Lindsay,~R.~K.; Buchanan,~B.~G.; Feigenbaum,~E.~A.; Lederberg,~J. DENDRAL: a
  case study of the first expert system for scientific hypothesis formation.
  \emph{Artificial intelligence} \textbf{1993}, \emph{61}, 209--261\relax
\mciteBstWouldAddEndPuncttrue
\mciteSetBstMidEndSepPunct{\mcitedefaultmidpunct}
{\mcitedefaultendpunct}{\mcitedefaultseppunct}\relax
\EndOfBibitem
\bibitem[Eng \latin{et~al.}(1994)Eng, McCormack, and Yates]{Eng1994sequest}
Eng,~J.~K.; McCormack,~A.~L.; Yates,~J.~R. An approach to correlate tandem mass
  spectral data of peptides with amino acid sequences in a protein database.
  \emph{Journal of the American Society for Mass Spectrometry} \textbf{1994},
  \emph{5}, 976--989\relax
\mciteBstWouldAddEndPuncttrue
\mciteSetBstMidEndSepPunct{\mcitedefaultmidpunct}
{\mcitedefaultendpunct}{\mcitedefaultseppunct}\relax
\EndOfBibitem
\bibitem[Tran \latin{et~al.}(2017)Tran, Zhang, Xin, Shan, and
  Li]{Tran8247deepnovo}
Tran,~N.~H.; Zhang,~X.; Xin,~L.; Shan,~B.; Li,~M. De novo peptide sequencing by
  deep learning. \emph{Proceedings of the National Academy of Sciences}
  \textbf{2017}, \emph{114}, 8247--8252\relax
\mciteBstWouldAddEndPuncttrue
\mciteSetBstMidEndSepPunct{\mcitedefaultmidpunct}
{\mcitedefaultendpunct}{\mcitedefaultseppunct}\relax
\EndOfBibitem
\bibitem[{Schoenholz} \latin{et~al.}(2018){Schoenholz}, {Hackett}, {Deming},
  {Melamud}, {Jaitly}, {McAllister}, {O'Brien}, {Dahl}, {Bennett}, {Dai}, and
  {Koller}]{Schoenholtz2018supervision}
{Schoenholz},~S.~S.; {Hackett},~S.; {Deming},~L.; {Melamud},~E.; {Jaitly},~N.;
  {McAllister},~F.; {O'Brien},~J.; {Dahl},~G.; {Bennett},~B.; {Dai},~A.~M.;
  {Koller},~D. {Peptide-spectra matching from weak supervision}.
  \emph{arXiv:1808.06576} \textbf{2018}, \relax
\mciteBstWouldAddEndPunctfalse
\mciteSetBstMidEndSepPunct{\mcitedefaultmidpunct}
{}{\mcitedefaultseppunct}\relax
\EndOfBibitem
\bibitem[D{\"u}hrkop \latin{et~al.}(2015)D{\"u}hrkop, Shen, Meusel, Rousu, and
  B{\"o}cker]{duhrkop2015searching}
D{\"u}hrkop,~K.; Shen,~H.; Meusel,~M.; Rousu,~J.; B{\"o}cker,~S. Searching
  molecular structure databases with tandem mass spectra using CSI: FingerID.
  \emph{Proceedings of the National Academy of Sciences} \textbf{2015},
  \emph{112}, 12580--12585\relax
\mciteBstWouldAddEndPuncttrue
\mciteSetBstMidEndSepPunct{\mcitedefaultmidpunct}
{\mcitedefaultendpunct}{\mcitedefaultseppunct}\relax
\EndOfBibitem
\bibitem[Curry and Rumelhart(1990)Curry, and Rumelhart]{curry1990msnet}
Curry,~B.; Rumelhart,~D.~E. MSnet: A neural network which classifies mass
  spectra. \emph{Tetrahedron Computer Methodology} \textbf{1990}, \emph{3},
  213--237\relax
\mciteBstWouldAddEndPuncttrue
\mciteSetBstMidEndSepPunct{\mcitedefaultmidpunct}
{\mcitedefaultendpunct}{\mcitedefaultseppunct}\relax
\EndOfBibitem
\bibitem[Rabinowitz(2017)]{spec2smiles}
Rabinowitz,~T. Mass-Spectrometry-Prediction.
  https://github.com/terryrabinowitz/Mass-Spectrometry-Prediction/blob/master/readme.pdf,
  2017; [Online; accessed 06-Oct-2018]\relax
\mciteBstWouldAddEndPuncttrue
\mciteSetBstMidEndSepPunct{\mcitedefaultmidpunct}
{\mcitedefaultendpunct}{\mcitedefaultseppunct}\relax
\EndOfBibitem
\bibitem[{Lim} \latin{et~al.}(2018){Lim}, {Wong}, {Wong}, {Tan}, {Chieu},
  {Choo}, and {Neo}]{Lim2018ChemicalStructure}
{Lim},~J.; {Wong},~J.; {Wong},~M.~X.; {Tan},~L. H.~E.; {Chieu},~H.~L.;
  {Choo},~D.; {Neo},~N. K.~N. {Chemical structure elucidation from mass
  spectrometry by matching substructures}. \emph{arXiv:1811.07886}
  \textbf{2018}, \relax
\mciteBstWouldAddEndPunctfalse
\mciteSetBstMidEndSepPunct{\mcitedefaultmidpunct}
{}{\mcitedefaultseppunct}\relax
\EndOfBibitem
\bibitem[Lorquet(1994)]{lorquet1994whither}
Lorquet,~J. Whither the statistical theory of mass spectra? \emph{Mass
  Spectrometry Reviews} \textbf{1994}, \emph{13}, 233--257\relax
\mciteBstWouldAddEndPuncttrue
\mciteSetBstMidEndSepPunct{\mcitedefaultmidpunct}
{\mcitedefaultendpunct}{\mcitedefaultseppunct}\relax
\EndOfBibitem
\bibitem[Lorquet(2000)]{lorquet2000landmarks}
Lorquet,~J.-C. Landmarks in the theory of mass spectra. \emph{International
  Journal of Mass Spectrometry} \textbf{2000}, \emph{200}, 43--56\relax
\mciteBstWouldAddEndPuncttrue
\mciteSetBstMidEndSepPunct{\mcitedefaultmidpunct}
{\mcitedefaultendpunct}{\mcitedefaultseppunct}\relax
\EndOfBibitem
\bibitem[Rosenstock \latin{et~al.}(1952)Rosenstock, Wallenstein, Wahrhaftig,
  and Eyring]{rosenstock1952absolute}
Rosenstock,~H.~M.; Wallenstein,~M.; Wahrhaftig,~A.; Eyring,~H. Absolute rate
  theory for isolated systems and the mass spectra of polyatomic molecules.
  \emph{Proceedings of the National Academy of Sciences} \textbf{1952},
  \emph{38}, 667--678\relax
\mciteBstWouldAddEndPuncttrue
\mciteSetBstMidEndSepPunct{\mcitedefaultmidpunct}
{\mcitedefaultendpunct}{\mcitedefaultseppunct}\relax
\EndOfBibitem
\bibitem[Irikura(2017)]{irikura2017ab}
Irikura,~K.~K. Ab initio computation of energy deposition during electron
  ionization of molecules. \emph{The Journal of Physical Chemistry A}
  \textbf{2017}, \emph{121}, 7751--7760\relax
\mciteBstWouldAddEndPuncttrue
\mciteSetBstMidEndSepPunct{\mcitedefaultmidpunct}
{\mcitedefaultendpunct}{\mcitedefaultseppunct}\relax
\EndOfBibitem
\bibitem[Ásgeirsson \latin{et~al.}(2017)Ásgeirsson, Bauer, and
  Grimme]{Asgeirsson_QCEIMS}
Ásgeirsson,~V.; Bauer,~C.~A.; Grimme,~S. Quantum chemical calculation of
  electron ionization mass spectra for general organic and inorganic molecules.
  \emph{Chem. Sci.} \textbf{2017}, \emph{8}, 4879--4895\relax
\mciteBstWouldAddEndPuncttrue
\mciteSetBstMidEndSepPunct{\mcitedefaultmidpunct}
{\mcitedefaultendpunct}{\mcitedefaultseppunct}\relax
\EndOfBibitem
\bibitem[Linstrom and Mallard(2001)Linstrom, and Mallard]{NIST_WebBook}
Linstrom,~P.~J.; Mallard,~W.~G. The NIST Chemistry WebBook: A chemical data
  resource on the internet. \emph{Journal of Chemical \& Engineering Data}
  \textbf{2001}, \emph{46}, 1059--1063\relax
\mciteBstWouldAddEndPuncttrue
\mciteSetBstMidEndSepPunct{\mcitedefaultmidpunct}
{\mcitedefaultendpunct}{\mcitedefaultseppunct}\relax
\EndOfBibitem
\bibitem[McLafferty \latin{et~al.}(1974)McLafferty, Hertel, and
  Villwock]{mclafferty1974probability}
McLafferty,~F.; Hertel,~R.; Villwock,~R. Probability based matching of mass
  spectra. Rapid identification of specific compounds in mixtures.
  \emph{Journal of Mass Spectrometry} \textbf{1974}, \emph{9}, 690--702\relax
\mciteBstWouldAddEndPuncttrue
\mciteSetBstMidEndSepPunct{\mcitedefaultmidpunct}
{\mcitedefaultendpunct}{\mcitedefaultseppunct}\relax
\EndOfBibitem
\bibitem[Hertz \latin{et~al.}(1971)Hertz, Hites, and
  Biemann]{hertz1971identification}
Hertz,~H.~S.; Hites,~R.~A.; Biemann,~K. Identification of mass spectra by
  computer-searching a file of known spectra. \emph{Analytical Chemistry}
  \textbf{1971}, \emph{43}, 681--691\relax
\mciteBstWouldAddEndPuncttrue
\mciteSetBstMidEndSepPunct{\mcitedefaultmidpunct}
{\mcitedefaultendpunct}{\mcitedefaultseppunct}\relax
\EndOfBibitem
\bibitem[Moorthy \latin{et~al.}(2017)Moorthy, Wallace, Kearsley, Tchekhovskoi,
  and Stein]{moorthy2017combining}
Moorthy,~A.~S.; Wallace,~W.~E.; Kearsley,~A.~J.; Tchekhovskoi,~D.~V.;
  Stein,~S.~E. Combining fragment-ion and neutral-loss matching during mass
  spectral library searching: A new general purpose algorithm applicable to
  illicit drug identification. \emph{Analytical chemistry} \textbf{2017},
  \emph{89}, 13261--13268\relax
\mciteBstWouldAddEndPuncttrue
\mciteSetBstMidEndSepPunct{\mcitedefaultmidpunct}
{\mcitedefaultendpunct}{\mcitedefaultseppunct}\relax
\EndOfBibitem
\bibitem[RDKit, online()]{rdkit}
{RDK}it: Open-source cheminformatics. \url{http://www.rdkit.org}, [Online;
  accessed 23-Sept-2018]\relax
\mciteBstWouldAddEndPuncttrue
\mciteSetBstMidEndSepPunct{\mcitedefaultmidpunct}
{\mcitedefaultendpunct}{\mcitedefaultseppunct}\relax
\EndOfBibitem
\bibitem[Rogers and Hahn(2010)Rogers, and Hahn]{Rogers_2010_ECFP}
Rogers,~D.; Hahn,~M. Extended-connectivity fingerprints. \emph{Journal of
  Chemical Information and Modeling} \textbf{2010}, \emph{50}, 742--754, PMID:
  20426451\relax
\mciteBstWouldAddEndPuncttrue
\mciteSetBstMidEndSepPunct{\mcitedefaultmidpunct}
{\mcitedefaultendpunct}{\mcitedefaultseppunct}\relax
\EndOfBibitem
\bibitem[Kingma and Ba(2014)Kingma, and Ba]{Kingma_adam_optimizer}
Kingma,~D.; Ba,~J. {Adam: A method for stochastic optimization}.
  \emph{arXiv:1412.6980} \textbf{2014}, \relax
\mciteBstWouldAddEndPunctfalse
\mciteSetBstMidEndSepPunct{\mcitedefaultmidpunct}
{}{\mcitedefaultseppunct}\relax
\EndOfBibitem
\bibitem[Abadi \latin{et~al.}(2016)Abadi, Barham, Chen, Chen, Davis, Dean,
  Devin, Ghemawat, Irving, Isard, Kudlur, Levenberg, Monga, Moore, Murray,
  Steiner, Tucker, Vasudevan, Warden, Wicke, Yu, and Zheng]{Tensorflow-2016}
Abadi,~M. \latin{et~al.}  TensorFlow: A system for large-scale machine
  learning. 12th USENIX Symposium on Operating Systems Design and
  Implementation (OSDI 16). 2016; pp 265--283\relax
\mciteBstWouldAddEndPuncttrue
\mciteSetBstMidEndSepPunct{\mcitedefaultmidpunct}
{\mcitedefaultendpunct}{\mcitedefaultseppunct}\relax
\EndOfBibitem
\bibitem[Golovin \latin{et~al.}(2017)Golovin, Solnik, Moitra, Kochanski, Karro,
  and Sculley]{Google_Vizier}
Golovin,~D.; Solnik,~B.; Moitra,~S.; Kochanski,~G.; Karro,~J.; Sculley,~D.
  Google vizier: A service for black-box optimization. Proceedings of the 23rd
  ACM SIGKDD International Conference on Knowledge Discovery and Data Mining.
  2017; pp 1487--1495\relax
\mciteBstWouldAddEndPuncttrue
\mciteSetBstMidEndSepPunct{\mcitedefaultmidpunct}
{\mcitedefaultendpunct}{\mcitedefaultseppunct}\relax
\EndOfBibitem
\bibitem[{He} \latin{et~al.}(2015){He}, {Zhang}, {Ren}, and {Sun}]{he_resnet}
{He},~K.; {Zhang},~X.; {Ren},~S.; {Sun},~J. {Deep residual learning for image
  recognition}. \emph{arXiv:1512.0338} \textbf{2015}, \relax
\mciteBstWouldAddEndPunctfalse
\mciteSetBstMidEndSepPunct{\mcitedefaultmidpunct}
{}{\mcitedefaultseppunct}\relax
\EndOfBibitem
\bibitem[Duvenaud \latin{et~al.}(2015)Duvenaud, Maclaurin, Iparraguirre,
  Bombarell, Hirzel, Aspuru-Guzik, and Adams]{duvenaud_convolutional_2015}
Duvenaud,~D.~K.; Maclaurin,~D.; Iparraguirre,~J.; Bombarell,~R.; Hirzel,~T.;
  Aspuru-Guzik,~A.; Adams,~R.~P. Convolutional networks on graphs for learning
  molecular fingerprints. \textbf{2015}, 2224--2232\relax
\mciteBstWouldAddEndPuncttrue
\mciteSetBstMidEndSepPunct{\mcitedefaultmidpunct}
{\mcitedefaultendpunct}{\mcitedefaultseppunct}\relax
\EndOfBibitem
\bibitem[{Gilmer} \latin{et~al.}(2017){Gilmer}, {Schoenholz}, {Riley},
  {Vinyals}, and {Dahl}]{gilmer_2017_mpnn}
{Gilmer},~J.; {Schoenholz},~S.~S.; {Riley},~P.~F.; {Vinyals},~O.; {Dahl},~G.~E.
  {Neural message passing for quantum chemistry}. \emph{arXiv:1704.0121}
  \textbf{2017}, \relax
\mciteBstWouldAddEndPunctfalse
\mciteSetBstMidEndSepPunct{\mcitedefaultmidpunct}
{}{\mcitedefaultseppunct}\relax
\EndOfBibitem
\bibitem[Kearnes \latin{et~al.}(2016)Kearnes, McCloskey, Berndl, Pande, and
  Riley]{kearnes2016molecular}
Kearnes,~S.; McCloskey,~K.; Berndl,~M.; Pande,~V.; Riley,~P. Molecular graph
  convolutions: moving beyond fingerprints. \emph{Journal of computer-aided
  molecular design} \textbf{2016}, \emph{30}, 595--608\relax
\mciteBstWouldAddEndPuncttrue
\mciteSetBstMidEndSepPunct{\mcitedefaultmidpunct}
{\mcitedefaultendpunct}{\mcitedefaultseppunct}\relax
\EndOfBibitem
\bibitem[{Landrum}(2014)]{rdkit_blogpost_collide_bits}
{Landrum},~G. Collding bits. 2014;
  \url{http://rdkit.blogspot.com/2014/02/colliding-bits.html}, [Online;
  accessed 11-Nov-2019]\relax
\mciteBstWouldAddEndPuncttrue
\mciteSetBstMidEndSepPunct{\mcitedefaultmidpunct}
{\mcitedefaultendpunct}{\mcitedefaultseppunct}\relax
\EndOfBibitem
\end{mcitethebibliography}

\end{document}